\documentclass[12pt]{article}

\usepackage[utf8]{inputenc}
\usepackage[T1]{fontenc}
\usepackage{amsmath,amssymb,amsthm}
\usepackage{graphicx}
\usepackage{booktabs}
\usepackage{natbib}
\usepackage[colorlinks=true,linkcolor=blue,citecolor=blue,urlcolor=blue]{hyperref}
\usepackage{siunitx}
\usepackage{caption}
\usepackage{subcaption}
\usepackage{geometry}
\usepackage{setspace}

\newcommand{\SMC}{S_{\text{MC}}}
\newcommand{\STV}{S_{\text{TV}}}

\newcommand{\TE}{\text{TE}}
\newcommand{\CTE}{\text{CTE}}
\newcommand{\II}{\text{II}}

\newcommand{\numNStocks}{100}  
\newcommand{\numNDates}{1,231}  
\newcommand{\numNObs}{123,100}  
\newcommand{\numEdgesFor}{45}  
\newcommand{\numDensityFor}{0.0045}  
\newcommand{\numMeanTEFor}{0.030}  
\newcommand{\numMaxTEFor}{0.069}  
\newcommand{\numEdgesInst}{50}  
\newcommand{\numDensityInst}{0.0051}  
\newcommand{\numMeanTEInst}{0.025}  
\newcommand{\numEdgesInd}{79}  
\newcommand{\numDensityInd}{0.0080}  
\newcommand{\numMeanTEInd}{0.015}  
\newcommand{\numMWpForInst}{0.278}  
\newcommand{\numIIMeanForInst}{-0.0018}  
\newcommand{\numIISynPctForInst}{1}  
\newcommand{\numIIMeanForInd}{-0.0026}  
\newcommand{\numIISynPctForInd}{3}  
\newcommand{\numIIpForInd}{0.018}  
\newcommand{\numIIMeanInstInd}{-0.0011}  
\newcommand{\numIISynPctInstInd}{1}  
\newcommand{\numIIWithinInst}{0.002}  
\newcommand{\numCTEForInstCIlo}{-0.0008}  
\newcommand{\numCTEForInstCIhi}{0.0082}  
\newcommand{\numCTEForIndCIlo}{-0.0008}  
\newcommand{\numCTEForIndCIhi}{0.0037}  
\newcommand{\numCTEInstIndCIlo}{-0.0121}  
\newcommand{\numCTEInstIndCIhi}{0.0041}  
\newcommand{\numCTEPerStockNForInst}{30}  
\newcommand{\numMIBits}{0.000}  
\newcommand{\numFanoBound}{52.2}  
\newcommand{\numCondEntropy}{0.692}  
\newcommand{\numRiskFreeRate}{0.035}  
\newcommand{\numFMbThreeForSTV}{0.191}  
\newcommand{\numFMtThreeForSTV}{2.04}  
\newcommand{\numFMpThreeForSTV}{0.042}  
\newcommand{\numFMbOneForSMC}{0.037}  
\newcommand{\numFMtOneForSMC}{0.41}  
\newcommand{\numFMbTwoForSMC}{-0.007}  
\newcommand{\numFMtTwoForSMC}{-0.54}  
\newcommand{\numFMbThreeForSMC}{-13.497}  
\newcommand{\numFMtThreeForSMC}{-0.86}  
\newcommand{\numFMbOneForSTV}{-0.003}  
\newcommand{\numFMtOneForSTV}{-3.79}  
\newcommand{\numFMbTwoForSTV}{0.005}  
\newcommand{\numFMtTwoForSTV}{0.38}  
\newcommand{\numFMbOneInstSMC}{0.468}  
\newcommand{\numFMtOneInstSMC}{3.29}  
\newcommand{\numFMbTwoInstSMC}{0.002}  
\newcommand{\numFMtTwoInstSMC}{0.14}  
\newcommand{\numFMbThreeInstSMC}{-17.195}  
\newcommand{\numFMtThreeInstSMC}{-1.17}  
\newcommand{\numFMbOneInstSTV}{0.004}  
\newcommand{\numFMtOneInstSTV}{4.14}  
\newcommand{\numFMbTwoInstSTV}{0.017}  
\newcommand{\numFMtTwoInstSTV}{1.39}  
\newcommand{\numFMbThreeInstSTV}{0.115}  
\newcommand{\numFMtThreeInstSTV}{1.12}  
\newcommand{\numFMbOneIndSMC}{-0.160}  
\newcommand{\numFMtOneIndSMC}{-1.77}  
\newcommand{\numFMbTwoIndSMC}{-0.016}  
\newcommand{\numFMtTwoIndSMC}{-1.58}  
\newcommand{\numFMbThreeIndSMC}{2.066}  
\newcommand{\numFMtThreeIndSMC}{0.25}  
\newcommand{\numFMbOneIndSTV}{0.000}  
\newcommand{\numFMtOneIndSTV}{0.57}  
\newcommand{\numFMbTwoIndSTV}{-0.002}  
\newcommand{\numFMtTwoIndSTV}{-0.17}  
\newcommand{\numFMbThreeIndSTV}{-0.038}  
\newcommand{\numFMtThreeIndSTV}{-0.62}  
\newcommand{\numFMnPeriods}{1231}  
\newcommand{\numPortImprovForSMC}{9.1 \times 10^{-7}}  
\newcommand{\numPortImprovIndSTV}{1.1 \times 10^{-4}}  
\newcommand{\numLagTEminFor}{0.0169}  
\newcommand{\numLagTEmaxFor}{0.0171}  
\newcommand{\numLagTEminInst}{0.0173}  
\newcommand{\numLagTEmaxInst}{0.0176}  
\newcommand{\numLagTEminInd}{0.0172}  
\newcommand{\numLagTEmaxInd}{0.0175}  
\newcommand{\numSymbolicTEFor}{0.015}  
\newcommand{\numKSGTEFor}{0.009}  
\newcommand{\numKSGTEInst}{0.008}  
\newcommand{\numThreshEdges}{5}  

\title{Information Propagation Across Investor Types: Transfer Entropy Networks in the Korean Equity Market}

\author{
    Sungwoo Kang\thanks{krml919@korea.ac.kr} \\
    Department of Electrical and Computer Engineering, Korea University \\
    Seoul 02841, Republic of Korea
}

\date{\today}

\begin{document}

\maketitle

\begin{abstract}
Whether heterogeneous investor flows transmit private information across stocks or merely reflect coordinated responses to public signals remains an open question in market microstructure. We construct Transfer Entropy (TE) networks from investor-type flows---foreign, institutional, and individual---for \numNStocks{} Korean equities over \numNDates{} trading days (January 2020 to February 2025), and evaluate their economic content through interaction information (II), conditional TE, mutual information (MI), Kelly criterion bounds, and Fama-MacBeth regressions. Three findings emerge. First, TE networks are sparse and structurally heterogeneous: foreign investors maintain few but strong links (\numEdgesFor{} edges, mean TE = \numMeanTEFor{}), while individual investors form many but weak links (\numEdgesInd{} edges, mean TE = \numMeanTEInd{}). Second, cross-investor information is redundant rather than synergistic, no investor type directionally dominates another, and MI between signals and returns is zero at the daily horizon. Third, network centrality adds negligible alpha in cross-sectional regressions, with only one of six signal-centrality interactions reaching marginal significance. These results indicate that the observed propagation structure captures shared information processing rather than private signal cascades, consistent with daily-frequency market efficiency.
\end{abstract}

\section{Introduction}

Information propagation across heterogeneous investor classes is central to understanding how financial markets aggregate dispersed beliefs into prices. The mechanism by which one investor type's trading activity influences another---whether through direct information transmission or through correlated responses to common factors---determines the extent to which cross-investor flow data contains exploitable signals for portfolio construction.

The Korea Exchange (KRX) provides a unique laboratory for studying this question. Korean securities regulation mandates daily disclosure of net trading volume by investor type---foreign, institutional, and individual---creating a public record of disaggregated order flow that is unavailable in most other markets. A substantial literature has exploited this disclosure regime to study relative performance and information content across investor classes. \citet{choe1999foreign} and \citet{choe2005domestic} established that foreign investors do not consistently destabilize Korean equity prices and that domestic investors may possess informational advantages in certain settings. More recently, \citet{kang2025matched} developed matched-filter signal definitions ($\SMC$ and $\STV$) that decompose investor-type flows into components aligned with market capitalization and trading volume, enabling signal-level analysis of information content.

Despite these advances, the literature lacks a systematic characterization of how investor-type information propagates across stocks. Existing studies treat each stock in isolation, estimating the predictive content of investor flows for that stock's own returns. Whether foreign investor activity in one stock predicts institutional or individual activity in another stock---and whether such cross-stock, cross-investor dependencies carry economic value---remains unexplored. Transfer Entropy (TE), an information-theoretic measure of directed statistical dependence introduced by \citet{schreiber2000measuring}, provides a natural framework for constructing such propagation networks. Unlike Granger causality \citep{granger1969investigating}, TE captures nonlinear dependencies without parametric assumptions, making it well suited to the potentially complex interactions among heterogeneous investor flows.

The gap matters because network topology is often assumed to proxy for exploitable structure. If TE networks reveal that certain stocks serve as bellwethers---transmitting information that predicts activity elsewhere---then centrality in these networks could constitute an alpha source. Conversely, if the network structure reflects only shared responses to common shocks, centrality would be uninformative for cross-sectional return prediction. Distinguishing between these scenarios has direct implications for the design of network-based trading strategies and for our understanding of market efficiency at the investor-type level.

We address this question through a seven-part empirical framework applied to \numNStocks{} stocks over \numNDates{} trading days. We construct TE networks for each investor type using symbolic TE with quantile-based discretization, assess statistical significance through block-permutation surrogates with false discovery rate (FDR) control, and characterize network topology through standard centrality measures. We then deploy higher-order information measures---interaction information and conditional TE---to test whether cross-investor flows carry synergistic or redundant information and whether any investor type directionally dominates another. Finally, we evaluate economic content through mutual information with returns, Kelly criterion growth rate bounds, and Fama-MacBeth cross-sectional regressions with signal-centrality interactions.

Three contributions emerge from this analysis. The first is a characterization of structural heterogeneity across investor-type TE networks. Foreign investor networks are sparse but carry high edge weights (\numEdgesFor{} significant edges at mean TE = \numMeanTEFor{}), institutional networks occupy an intermediate position (\numEdgesInst{} edges at mean TE = \numMeanTEInst{}), and individual investor networks are denser but weaker (\numEdgesInd{} edges at mean TE = \numMeanTEInd{}). Mann-Whitney tests confirm that edge weight distributions differ significantly between foreign and individual networks ($p < 0.001$) but not between foreign and institutional networks ($p = \numMWpForInst{}$). This inverse relationship between connectivity and edge strength suggests fundamentally different information processing architectures across investor types.

The second contribution is a comprehensive demonstration that this structural heterogeneity does not translate into informational asymmetry. Interaction information is negative for all cross-investor pairs, indicating that the information carried by any two investor types about returns is redundant rather than synergistic. Conditional TE is exactly zero at the aggregate level for all investor-pair comparisons, ruling out directional information dominance. Mutual information between investor-type signals and next-day returns is zero, implying that the Kelly-optimal growth rate equals the risk-free rate and that the Fano inequality bounds directional accuracy at \numFanoBound{}\%. These convergent null results establish that the TE network structure captures shared information processing rather than private signal cascades.

The third contribution is a direct test of the economic value of network topology. Fama-MacBeth regressions that interact signal strength with network centrality yield only one marginally significant coefficient out of six specifications (foreign $\STV$ signal-centrality interaction, $t = \numFMtThreeForSTV{}$), with no monotonic pattern across centrality quintiles. Network-enhanced portfolios improve annualized returns by less than 0.01 percentage points relative to signal-only portfolios. These findings confirm the structure-value disconnect: TE networks are statistically well-defined objects that carry no incremental economic information at the daily frequency.

The remainder of this paper proceeds as follows: Section~\ref{sec:literature} reviews related work, Section~\ref{sec:methodology} describes data and methods, Section~\ref{sec:results} presents results, Section~\ref{sec:discussion} discusses implications, and Section~\ref{sec:conclusion} concludes.

\section{Literature Review}
\label{sec:literature}

\subsection{Investor Heterogeneity in Financial Markets}

The Korean equity market occupies a distinctive position in the study of investor heterogeneity due to its mandatory disclosure of trading activity by investor type. The KRX classifies all market participants into three categories---foreign, institutional, and individual---and publishes daily net buy and sell volumes for each category at the individual stock level. This transparency has generated a substantial empirical literature on the relative information content and performance of each investor class.

\citet{choe1999foreign} examined whether foreign investors destabilized the Korean market during the 1997 Asian financial crisis. Their evidence suggested that foreign investors exhibited positive feedback trading but did not systematically destabilize prices, as the price impact of their trades was largely temporary. \citet{choe2005domestic} extended this analysis to normal market conditions and found that domestic institutional investors traded with somewhat better information than foreign investors, particularly for smaller and less liquid stocks where local knowledge confers advantages. These findings established that investor-type classification carries informational content, though the direction and magnitude of any informational advantage depend on market conditions and stock characteristics.

More recent work has refined the signal definitions used to extract information from investor-type flows. \citet{kang2025matched} introduced matched-filter signals ($\SMC$ and $\STV$) that decompose net investor flows into components correlated with market capitalization rank and trading volume, respectively. The matched-filter approach treats investor flows as noisy observations of latent information signals and applies signal processing techniques to extract the component most relevant for return prediction. This paper builds directly on that signal framework, using $\SMC$ and $\STV$ as the base signals whose cross-stock propagation properties we investigate through TE networks.

\subsection{Information-Theoretic Approaches in Finance}

Information theory provides a model-free framework for quantifying statistical dependencies that complements and extends traditional regression-based approaches. \citet{shannon1948mathematical} established the foundational measures---entropy, mutual information, and channel capacity---that quantify information content and transmission without parametric assumptions. \citet{cover2006elements} developed the connections between information theory and gambling that underpin the Kelly criterion framework used in this paper.

Transfer Entropy, introduced by \citet{schreiber2000measuring}, extends mutual information to capture directed, time-lagged dependencies between stochastic processes. Unlike Granger causality \citep{granger1969investigating}, which assumes linear dynamics and Gaussian innovations, TE measures the reduction in uncertainty about the future of a target process achieved by conditioning on the past of a source process, beyond what is already explained by the target's own history. This makes TE particularly suitable for financial applications where nonlinear dependencies and non-Gaussian distributions are prevalent.

\citet{dimpfl2013using} applied TE to measure information flows between the S\&P~500 and DAX index futures, demonstrating that TE captures lead-lag relationships that Granger causality misses during periods of market stress. \citet{sensoy2014effective} constructed TE-based information flow networks across exchange rates and stock markets, finding that the topology of information transmission shifts substantially during crisis periods. These applications established TE as a viable tool for financial network analysis, but they focused on market-level or asset-class-level transmission. The application of TE to investor-type flow networks within a single equity market---the focus of this paper---remains unexplored.

The higher-order information measures we employ extend the basic TE framework. Interaction information, formalized in the partial information decomposition framework of \citet{williams2010nonnegative}, quantifies whether two sources carry synergistic (complementary) or redundant (overlapping) information about a target variable. The Kraskov-St\"ogbauer-Grassberger (KSG) estimator \citep{kraskov2004estimating} provides a consistent, non-parametric method for estimating mutual information and related quantities from continuous data, which we use for conditional TE estimation.

\subsection{Network Approaches to Financial Markets}

Network methods have become standard tools for characterizing the structure of financial markets. \citet{mantegna1999hierarchical} pioneered the construction of minimum spanning trees from correlation matrices of stock returns, revealing hierarchical cluster structure that corresponds to economic sectors. This correlation-based approach has been widely extended to study portfolio diversification, systemic risk, and market taxonomy.

A key limitation of correlation networks is that they capture only undirected, contemporaneous associations. \citet{billio2012econometric} addressed this limitation by constructing directed networks from Granger causality relationships among financial institutions, providing measures of systemic risk and connectedness that proved relevant during the 2008 financial crisis. Their approach demonstrated that directed network topology carries information about systemic risk beyond what is captured by pairwise correlations.

The connection between network centrality and asset returns has been explored in several settings. Stocks that occupy central positions in correlation networks tend to have higher systematic risk, suggesting that centrality proxies for exposure to common factors rather than for private information. Whether centrality in information-theoretic (TE) networks---which capture directed, potentially nonlinear dependencies---carries different economic content from centrality in correlation networks is an open question that motivates our Fama-MacBeth analysis. The self-exciting dynamics that can arise in information networks have been modeled using Hawkes processes \citep{hawkes1971spectra}, and the microstructure foundations of information transmission have been formalized in the tradition of \citet{kyle1985continuous}. Our work connects these theoretical perspectives to the empirical TE network framework.

\section{Data and Methodology}
\label{sec:methodology}

\subsection{Data}

Our sample comprises daily investor-type trading data for \numNStocks{} Korean equities listed on the KRX, selected as the largest stocks by market capitalization with sufficient trading history. The sample spans \numNDates{} trading days from January 2020 through February 2025. For each stock on each day, we observe net buy volume decomposed by investor type (foreign, institutional, individual), closing price, market capitalization, and trading volume.

From the raw flow data, we construct two signal types following \citet{kang2025matched}. The market-capitalization signal $\SMC$ captures the component of net investor flow that co-moves with market capitalization rank, while the trading-volume signal $\STV$ captures the component co-moving with trading volume. Each investor type (foreign, institutional, individual) has a matched signal that the prior study identified as most informative: $\STV$ for foreign investors, $\SMC$ for institutional investors, and $\SMC$ for individual investors.

Table~\ref{tab:data_summary} summarizes the dataset. The sample contains \numNStocks{} stocks observed over \numNDates{} trading days, yielding \numNObs{} stock-day observations per investor type. Significant TE edges number \numEdgesFor{} for foreign, \numEdgesInst{} for institutional, and \numEdgesInd{} for individual investor networks, foreshadowing the structural heterogeneity that we characterize in detail below.

\begin{table}[htbp]
\centering
\caption{Data Summary: TE Network Dataset}
\label{tab:data_summary}
\small
\begin{tabular}{lrrr}
\toprule
Statistic & Foreign & Institutional & Individual \\
\midrule
N stocks & \multicolumn{3}{c}{100} \\
N trading days & \multicolumn{3}{c}{1231} \\
\addlinespace
Foreign -- N sig. edges & \multicolumn{3}{c}{45} \\
Institutional -- N sig. edges & \multicolumn{3}{c}{50} \\
Individual -- N sig. edges & \multicolumn{3}{c}{79} \\
\bottomrule
\end{tabular}
\end{table}

\subsection{Transfer Entropy Estimation}

Transfer Entropy from a source process $Y$ to a target process $X$ quantifies the reduction in uncertainty about the future of $X$ obtained by conditioning on the past of $Y$, beyond what is already explained by the past of $X$ itself. Formally,
\begin{equation}
\label{eq:te}
\TE(Y \to X) = \sum p(x_{t+1}, x_t^{(k)}, y_t^{(l)}) \log \frac{p(x_{t+1} | x_t^{(k)}, y_t^{(l)})}{p(x_{t+1} | x_t^{(k)})},
\end{equation}
where $x_t^{(k)}$ denotes the $k$-dimensional history of $X$ up to time $t$, $y_t^{(l)}$ denotes the $l$-dimensional history of $Y$, and the sum runs over all observed states \citep{schreiber2000measuring}. TE is non-negative and equals zero if and only if $Y$ provides no additional predictive information about $X$ beyond its own history.

We estimate TE using symbolic discretization. Continuous flow signals are mapped to five quantile-based symbols, converting the estimation problem from density estimation in continuous space to frequency counting in a discrete alphabet. We set the history lengths to $k = 1$ and $l = 1$, yielding a manageable state space of $5^3 = 125$ joint states while capturing one-step-ahead dependencies. The target variable for TE computation is the investor-type flow signal itself (flow-to-flow TE), measuring how one stock's investor flow predicts another stock's investor flow at the same investor-type level.

\subsection{Network Construction and Significance Testing}

For each investor type, we compute the pairwise TE matrix across all \numNStocks{} stocks, yielding $\numNStocks{} \times 99 = 9{,}900$ directed pairs. To identify statistically significant edges, we generate 200 block-permutation surrogates for each pair. Block permutation preserves the autocorrelation structure of the source series by permuting blocks of 20 consecutive observations rather than individual data points, providing a conservative null distribution that accounts for temporal dependence.

An edge is deemed significant if the observed TE exceeds the 95th percentile of the surrogate distribution. We apply the \citet{benjamini1995controlling} FDR correction at $\alpha = 0.05$ to control the expected proportion of false discoveries across the $9{,}900$ tests conducted for each investor type. The resulting binary adjacency matrices define the TE networks, and the significant TE values serve as edge weights. We characterize network topology through standard centrality measures: out-degree (number of outgoing edges), in-degree (number of incoming edges), betweenness centrality (fraction of shortest paths passing through a node), and closeness centrality (inverse average shortest path length).

\subsection{Higher-Order Information Measures}

To assess whether cross-investor information is synergistic or redundant, we compute the Interaction Information (II) among pairs of investor-type flow signals and stock returns. For two source variables $A$ and $B$ and a target $R$,
\begin{equation}
\label{eq:ii}
\II(A; B; R) = I(A, B; R) - I(A; R) - I(B; R),
\end{equation}
where $I(\cdot; \cdot)$ denotes mutual information and $I(A,B; R)$ is the joint mutual information of $(A, B)$ about $R$ \citep{williams2010nonnegative}. Positive II indicates synergy: the two sources together carry more information about the target than the sum of their individual contributions. Negative II indicates redundancy: the sources share overlapping information.

To test whether any investor type directionally dominates another in predicting returns, we compute Conditional Transfer Entropy (CTE). For investor-type signals $A$ and $B$ and return target $R$,
\begin{equation}
\label{eq:cte}
\CTE(A \to R \mid B) = \TE(A \to R) - \TE(A \to R \mid B),
\end{equation}
where $\TE(A \to R \mid B)$ conditions the TE computation on the history of $B$ using the KSG estimator \citep{kraskov2004estimating} with $k = 5$ neighbors. CTE measures how much of $A$'s predictive power for $R$ is unique---not shared with $B$.

From the CTE values, we construct a Directionality Index to summarize the relative information leadership between two investor types:
\begin{equation}
\label{eq:directionality}
D(A, B) = \CTE(A \to R \mid B) - \CTE(B \to R \mid A).
\end{equation}
A positive $D$ indicates that investor type $A$ provides more unique information about returns than $B$; $D = 0$ indicates symmetric information contributions. We estimate confidence intervals for $D$ through block bootstrap resampling.

\subsection{Kelly Criterion and Information Bounds}

To translate information-theoretic quantities into economic terms, we compute the mutual information between investor-type signals and next-day returns:
\begin{equation}
\label{eq:mi}
I(X; R) = H(R) - H(R \mid X),
\end{equation}
where $H(R)$ is the entropy of next-day returns and $H(R \mid X)$ is the conditional entropy given the signal $X$ \citep{cover2006elements}. MI quantifies the maximum information available for return prediction, measured in bits.

The Kelly criterion provides a direct mapping from MI to the maximum achievable growth rate. \citet{kelly1956new} showed that the optimal growth rate of a gambler with side information is
\begin{equation}
\label{eq:kelly}
G^* = r_f + I(X; R),
\end{equation}
where $r_f$ is the risk-free rate and $I(X; R)$ is the mutual information between the signal and the outcome, both expressed in compatible units. When $I(X; R) = 0$, the Kelly-optimal strategy degenerates to the risk-free asset, confirming that the signal carries no exploitable information.

The Fano inequality provides a complementary bound on directional accuracy. For a binary prediction problem (up or down),
\begin{equation}
\label{eq:fano}
P_e \geq \frac{H(R \mid X) - 1}{\log(|\mathcal{R}| - 1)},
\end{equation}
where $P_e$ is the minimum achievable error probability and $|\mathcal{R}|$ is the number of return categories \citep{fano1961transmission}. This inequality translates the conditional entropy $H(R \mid X)$ into a ceiling on the directional accuracy of any predictor based on signal $X$, regardless of the prediction algorithm employed.

\subsection{Cross-Sectional Regressions}

To test whether network centrality enhances the cross-sectional predictive power of investor-type signals, we estimate Fama-MacBeth regressions \citep{fama1973risk} of the form
\begin{equation}
\label{eq:fm}
R_{i,t+1} = a_t + b_{1,t} \cdot \text{Signal}_{i,t} + b_{2,t} \cdot \text{Centrality}_{i,t} + b_{3,t} \cdot \text{Signal}_{i,t} \times \text{Centrality}_{i,t} + \varepsilon_{i,t},
\end{equation}
where $R_{i,t+1}$ is the next-period return of stock $i$, $\text{Signal}_{i,t}$ is the investor-type flow signal ($\SMC$ or $\STV$), and $\text{Centrality}_{i,t}$ is the out-degree centrality of stock $i$ in the corresponding investor-type TE network. The interaction term $b_3$ is the coefficient of primary interest: a significant $b_3$ would indicate that network centrality amplifies or attenuates the signal's predictive power. We estimate Eq.~(\ref{eq:fm}) cross-sectionally at each of the \numNDates{} trading days and report time-series averages of the coefficient estimates with Fama-MacBeth $t$-statistics.

\section{Results}
\label{sec:results}

\subsection{Transfer Entropy Network Structure}

Table~\ref{tab:te_network_stats} reports the structural properties of TE networks constructed for each investor type. The foreign investor network contains \numEdgesFor{} significant edges with a density of \numDensityFor{}, a mean edge weight (TE) of \numMeanTEFor{}, and a maximum TE of \numMaxTEFor{}. The institutional investor network is moderately denser with \numEdgesInst{} edges (density \numDensityInst{}) and lower mean TE of \numMeanTEInst{}. The individual investor network is the densest with \numEdgesInd{} edges (density \numDensityInd{}) but has the weakest edges at mean TE of \numMeanTEInd{}.

\begin{table}[htbp]
\centering
\caption{Transfer Entropy Network Statistics}
\label{tab:te_network_stats}
\small
\begin{tabular}{lrrr}
\toprule
Metric & Foreign & Institutional & Individual \\
\midrule
N sig. edges & 45 & 50 & 79 \\
Density & 0.0045 & 0.0051 & 0.0080 \\
Mean TE & 0.030182 & 0.025328 & 0.015448 \\
Max TE & 0.068521 & 0.057162 & 0.048226 \\
Mean Clustering & 0.011667 & 0.000000 & 0.000000 \\
\addlinespace
Edge TE Mean & 0.030182 & 0.025328 & 0.015448 \\
Edge TE Median & 0.028248 & 0.026164 & 0.013584 \\
\addlinespace
MW $p$: Foreign Vs Institutional & \multicolumn{3}{c}{0.2781} \\
MW $p$: Foreign Vs Individual & \multicolumn{3}{c}{0.0000***} \\
MW $p$: Institutional Vs Individual & \multicolumn{3}{c}{0.0001***} \\
\bottomrule
\multicolumn{4}{l}{\footnotesize TE computed via symbolic method (ternary). MW = Mann-Whitney U test on edge weights.} \\
\multicolumn{4}{l}{\footnotesize $^{***}p<0.01$, $^{**}p<0.05$, $^{*}p<0.10$.} \\
\end{tabular}
\end{table}

This inverse relationship between edge count and edge strength is the defining structural feature of the data. Foreign investor TE networks are characterized by selective but intense information coupling: when foreign flows in one stock predict foreign flows in another, the predictive relationship is strong. Individual investor networks display the opposite pattern---diffuse but weak connections---consistent with correlated but low-precision responses to common signals such as media coverage or market-wide sentiment. Mann-Whitney tests on edge weight distributions confirm that foreign and individual networks differ significantly ($p < 0.001$), as do institutional and individual networks ($p < 0.001$), while foreign and institutional networks are statistically indistinguishable ($p = \numMWpForInst{}$).

\begin{figure}[htbp]
\centering
\includegraphics[width=\textwidth]{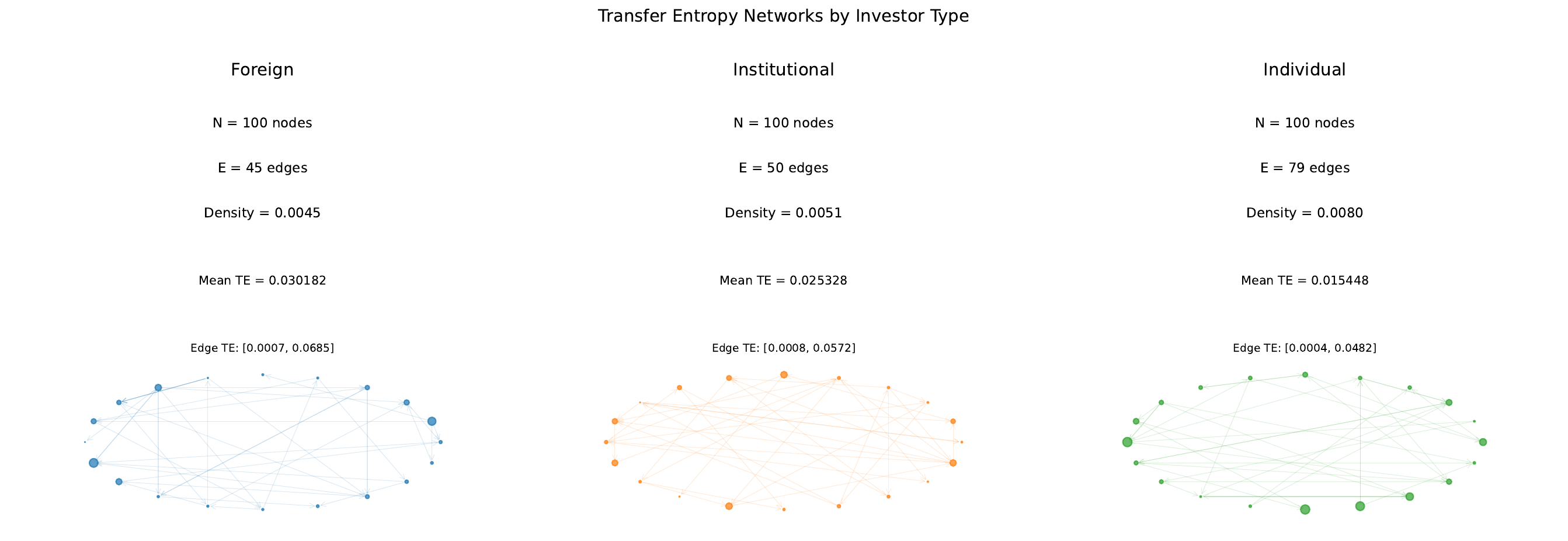}
\caption{Directed Transfer Entropy network graphs for foreign (left), institutional (center), and individual (right) investor types. Nodes represent the \numNStocks{} stocks; edges indicate statistically significant TE at FDR-corrected $\alpha = 0.05$ using 200 block-permutation surrogates. Edge width is proportional to TE magnitude. Foreign networks exhibit few but strong connections, while individual networks are denser but weaker.}
\label{fig:te_network}
\end{figure}

\begin{figure}[htbp]
\centering
\includegraphics[width=\textwidth]{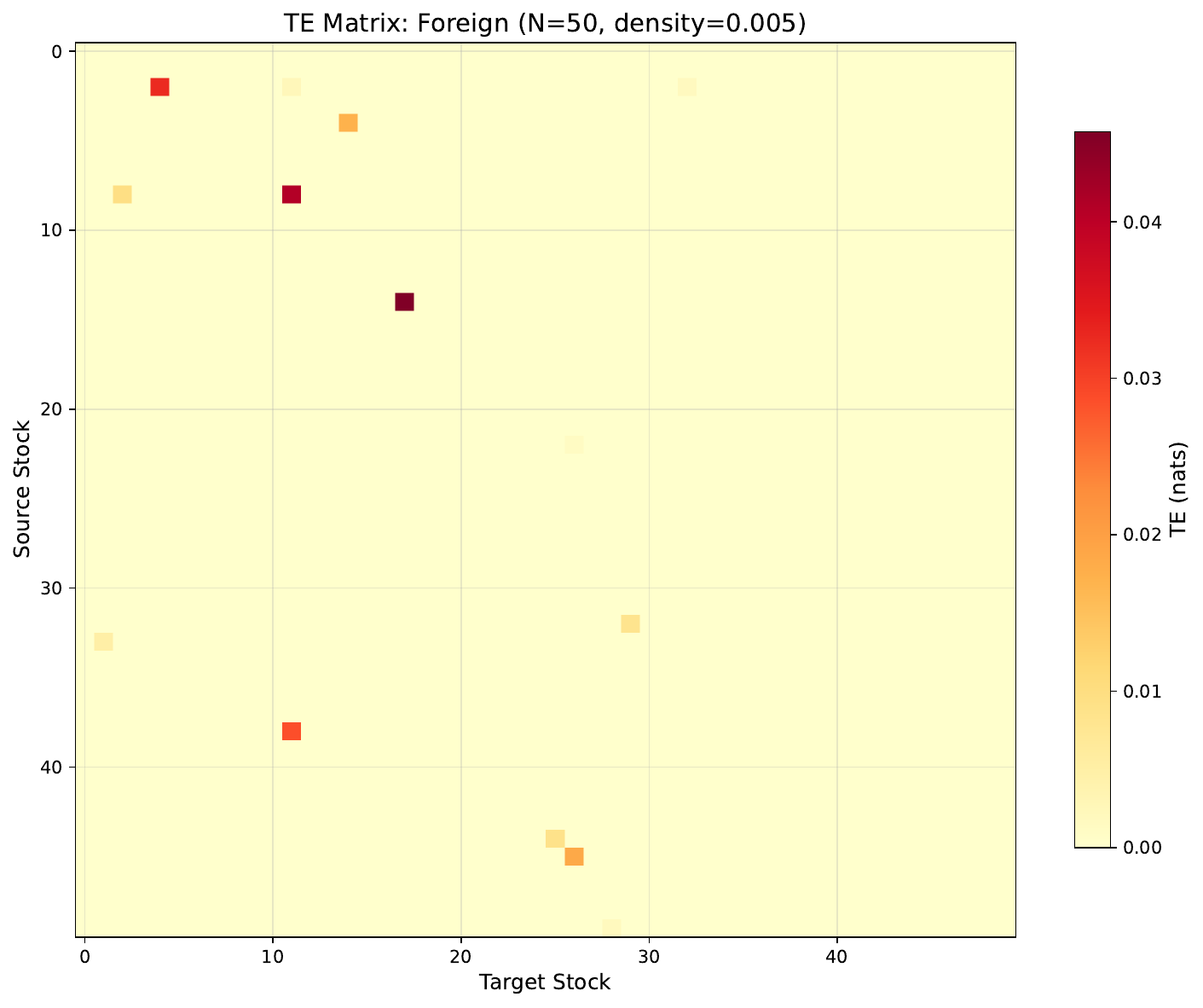}
\caption{Pairwise Transfer Entropy heatmaps for each investor type. Rows represent source stocks and columns represent target stocks. Color intensity indicates TE magnitude. The sparse structure confirms that significant information propagation is confined to a small subset of stock pairs across all investor types.}
\label{fig:te_heatmap}
\end{figure}

Figure~\ref{fig:te_network} visualizes the directed TE network graphs for each investor type. The foreign network exhibits a hub-and-spoke structure with a small number of high-out-degree nodes, whereas the individual network is more uniformly connected. Figure~\ref{fig:te_heatmap} displays the full pairwise TE matrices as heatmaps, revealing that most stock pairs have near-zero TE, with significant edges concentrated in a small fraction of pairs.

\subsection{Bellwether Stocks and Centrality}

Table~\ref{tab:bellwethers} ranks the top 10 bellwether stocks by out-degree centrality for each investor type. The composition of bellwether lists differs substantially across investor types. Only one stock (A042670) appears among the top 10 for both foreign and institutional networks, and only two stocks (A000660, A086520) appear for both foreign and individual networks. This lack of overlap indicates that the information propagation hubs differ by investor type, consistent with the structural heterogeneity documented in the previous subsection.

\begin{table}[htbp]
\centering
\caption{Top Bellwether Stocks by Out-Degree Centrality}
\label{tab:bellwethers}
\small
\begin{tabular}{clrrrr}
\toprule
Rank & Ticker & Out-Degree & Wtd Out-Deg & PageRank & Betweenness \\
\midrule
\multicolumn{6}{l}{\textbf{Foreign}} \\
1 & A009540 & 0.0303 & 0.069564 & 0.0092 & 0.0005 \\
2 & A047810 & 0.0202 & 0.065292 & 0.0062 & 0.0000 \\
3 & A010950 & 0.0202 & 0.043873 & 0.0213 & 0.0009 \\
4 & A008770 & 0.0202 & 0.064605 & 0.0115 & 0.0003 \\
5 & A041510 & 0.0202 & 0.047444 & 0.0062 & 0.0000 \\
6 & A042670 & 0.0202 & 0.059743 & 0.0062 & 0.0000 \\
7 & A000660 & 0.0101 & 0.014134 & 0.0062 & 0.0000 \\
8 & A035720 & 0.0101 & 0.036106 & 0.0062 & 0.0000 \\
9 & A373220 & 0.0101 & 0.046948 & 0.0062 & 0.0000 \\
10 & A096770 & 0.0101 & 0.019642 & 0.0168 & 0.0002 \\
\addlinespace
\multicolumn{6}{l}{\textbf{Institutional}} \\
1 & A028300 & 0.0303 & 0.089091 & 0.0141 & 0.0019 \\
2 & A326030 & 0.0303 & 0.069308 & 0.0067 & 0.0000 \\
3 & A086520 & 0.0202 & 0.011162 & 0.0238 & 0.0012 \\
4 & A086790 & 0.0202 & 0.086058 & 0.0090 & 0.0012 \\
5 & A011070 & 0.0202 & 0.056552 & 0.0184 & 0.0006 \\
6 & A003490 & 0.0202 & 0.081028 & 0.0152 & 0.0010 \\
7 & A316140 & 0.0202 & 0.068253 & 0.0067 & 0.0000 \\
8 & A011170 & 0.0202 & 0.011608 & 0.0395 & 0.0021 \\
9 & A004020 & 0.0202 & 0.073153 & 0.0124 & 0.0005 \\
10 & A042670 & 0.0202 & 0.079154 & 0.0067 & 0.0000 \\
\addlinespace
\multicolumn{6}{l}{\textbf{Individual}} \\
1 & A003550 & 0.0404 & 0.058008 & 0.0057 & 0.0000 \\
2 & A055550 & 0.0303 & 0.100223 & 0.0209 & 0.0047 \\
3 & A012330 & 0.0303 & 0.041655 & 0.0057 & 0.0000 \\
4 & A064350 & 0.0303 & 0.088808 & 0.0237 & 0.0067 \\
5 & A293490 & 0.0303 & 0.070996 & 0.0106 & 0.0005 \\
6 & A277810 & 0.0303 & 0.032021 & 0.0264 & 0.0080 \\
7 & A000660 & 0.0202 & 0.012275 & 0.0059 & 0.0006 \\
8 & A086520 & 0.0202 & 0.039522 & 0.0057 & 0.0000 \\
9 & A086790 & 0.0202 & 0.045275 & 0.0057 & 0.0000 \\
10 & A011790 & 0.0202 & 0.025316 & 0.0057 & 0.0000 \\
\addlinespace
\bottomrule
\end{tabular}
\end{table}

\begin{figure}[htbp]
\centering
\includegraphics[width=\textwidth]{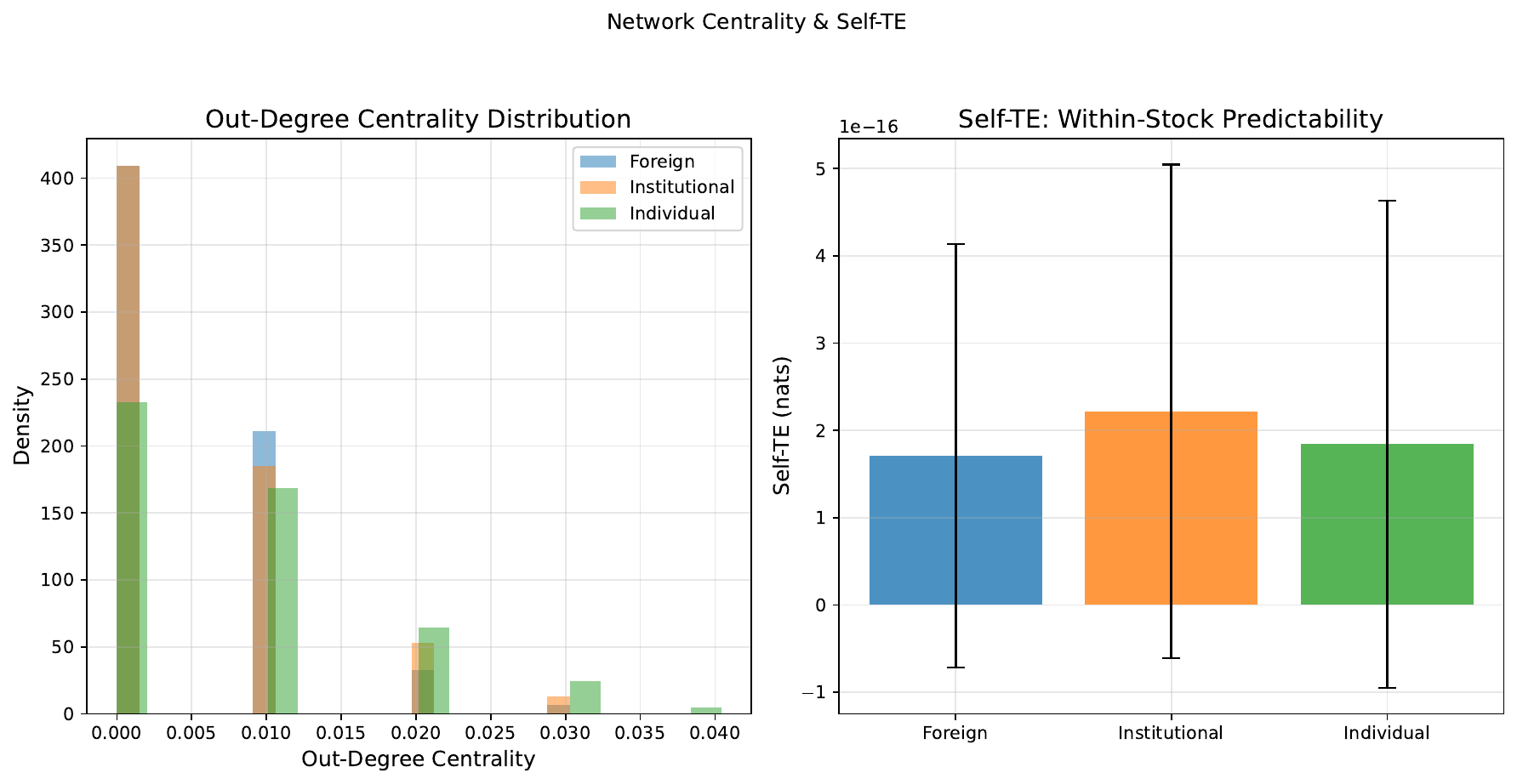}
\caption{Distributions of network centrality measures (out-degree, in-degree, betweenness, closeness) across stocks for each investor type. All distributions are right-skewed, with the foreign network exhibiting the highest concentration of centrality among a small number of hub stocks.}
\label{fig:centrality}
\end{figure}

Figure~\ref{fig:centrality} displays the distributions of out-degree, in-degree, betweenness, and closeness centrality across stocks for each investor type. All distributions are heavily right-skewed, with most stocks having zero or near-zero centrality and a small number of stocks occupying hub positions. The foreign network shows the most concentrated centrality distribution, reflecting its hub-and-spoke topology. The individual network shows a more uniform spread, consistent with its higher density and lower concentration of information flow.

\subsection{Interaction Information}

Table~\ref{tab:interaction_info} reports the interaction information analysis. All cross-investor pair comparisons yield negative mean II values: $\numIIMeanForInst{}$ for foreign versus institutional, $\numIIMeanForInd{}$ for foreign versus individual, and $\numIIMeanInstInd{}$ for institutional versus individual. The fraction of stocks exhibiting synergistic (positive) II is \numIISynPctForInst{}\%, \numIISynPctForInd{}\%, and \numIISynPctInstInd{}\%, respectively. Only the foreign-versus-individual comparison is statistically significant ($p = \numIIpForInd{}$); the remaining comparisons cannot reject $\II = 0$.

\begin{table}[htbp]
\centering
\caption{Interaction Information: Synergy vs Redundancy}
\label{tab:interaction_info}
\small
\begin{tabular}{lrrrrrr}
\toprule
Pair & Mean II & Median II & \% Synergy & MI(A;R) & MI(B;R) & $p$-value \\
\midrule
\multicolumn{7}{l}{\textit{Cross-Investor}} \\
Foreign Vs Institutional & -0.001805 & -0.003396 & 0.01 & 0.004966 & 0.007656 & 0.3969 \\
Foreign Vs Individual & -0.002560 & -0.003238 & 0.03 & 0.004966 & 0.005204 & 0.0182** \\
Institutional Vs Individual & -0.001063 & -0.003541 & 0.01 & 0.007656 & 0.005204 & 0.7119 \\
\addlinespace
\multicolumn{7}{l}{\textit{Within-Investor (S\_MC vs S\_TV)}} \\
Foreign & -0.000501 & -0.001663 & 0.21 & -- & -- & 0.7247 \\
Institutional & 0.002456 & -0.001990 & 0.13 & -- & -- & 0.6200 \\
Individual & -0.000621 & -0.002472 & 0.16 & -- & -- & 0.7361 \\
\bottomrule
\multicolumn{7}{l}{\footnotesize II > 0 = synergy (orthogonal info). II < 0 = redundancy (crowded). $p$-value from $t$-test of II $\neq$ 0.} \\
\multicolumn{7}{l}{\footnotesize $^{***}p<0.01$, $^{**}p<0.05$, $^{*}p<0.10$.} \\
\end{tabular}
\end{table}

The uniformly negative II values indicate that cross-investor flow signals carry redundant information about returns. When foreign and individual flow signals are combined, their joint information about returns is less than the sum of their individual contributions. This redundancy is inconsistent with models in which different investor types observe independent private signals and transmit them through trading, as such models predict positive II (synergy). Instead, the pattern is consistent with all investor types responding---with different intensities and noise levels---to the same underlying public information.

\begin{figure}[htbp]
\centering
\includegraphics[width=\textwidth]{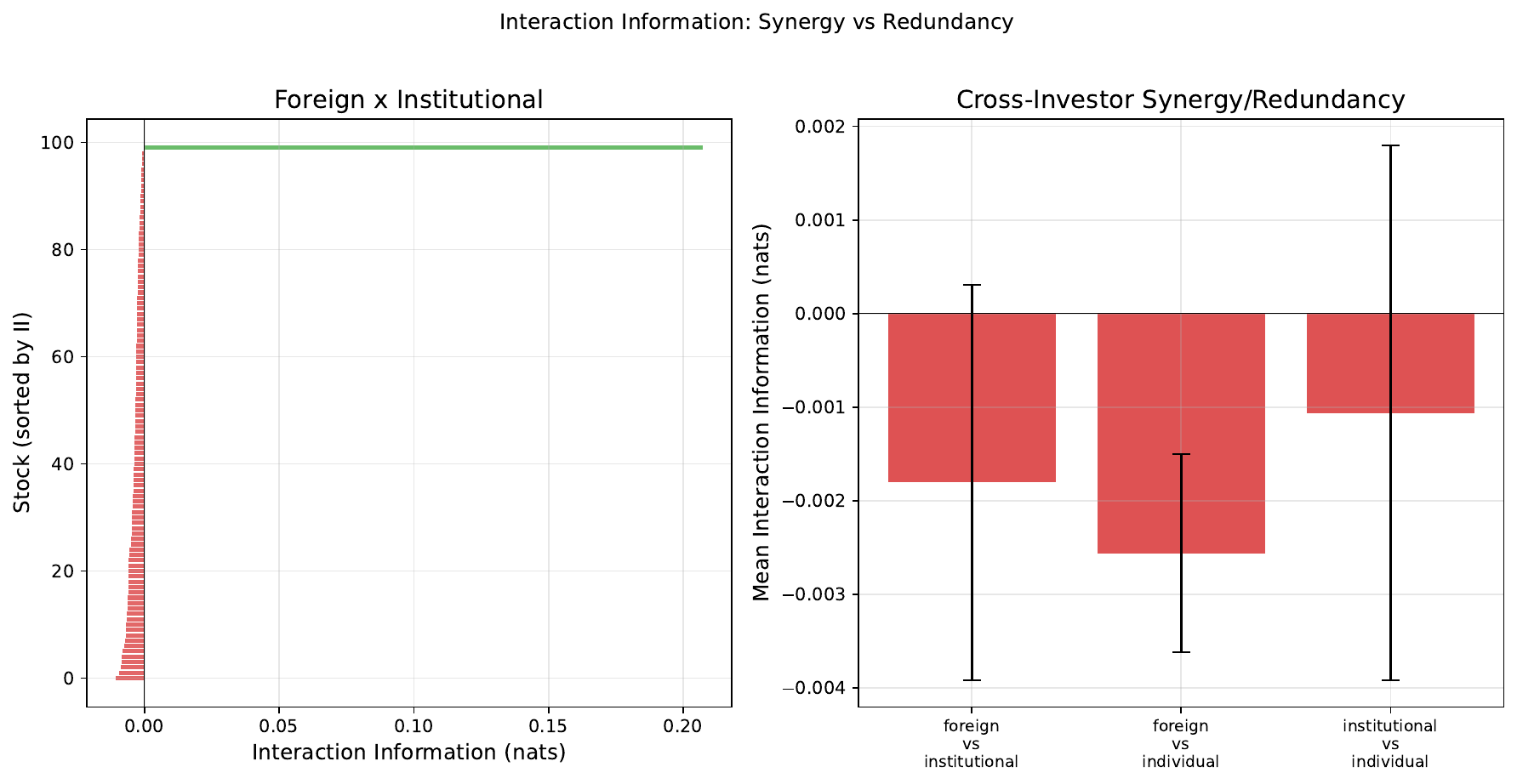}
\caption{Distribution of per-stock interaction information (II) values for each cross-investor pair. Negative values indicate redundancy; positive values indicate synergy. The distributions are centered slightly below zero, with nearly all stocks exhibiting redundant cross-investor information about returns.}
\label{fig:interaction}
\end{figure}

Figure~\ref{fig:interaction} visualizes the distribution of per-stock II values across the \numNStocks{} stocks for each investor pair. The distributions are centered near zero with slight negative skew, confirming that the aggregate results are not driven by a small number of outlier stocks.

\subsection{Conditional Transfer Entropy}

Table~\ref{tab:conditional_te} presents the conditional TE and directionality analysis. At the aggregate level, CTE is exactly zero for all three investor-pair comparisons. The directionality index $D$ is zero in each case, with bootstrap 95\% confidence intervals that include zero (foreign versus institutional: $[\numCTEForInstCIlo{}, \numCTEForInstCIhi{}]$; foreign versus individual: $[\numCTEForIndCIlo{}, \numCTEForIndCIhi{}]$; institutional versus individual: $[\numCTEInstIndCIlo{}, \numCTEInstIndCIhi{}]$).

\begin{table}[htbp]
\centering
\caption{Conditional Transfer Entropy and Directionality Index}
\label{tab:conditional_te}
\small
\begin{tabular}{lrrrrr}
\toprule
Pair & CTE(A$\to$R$|$B) & CTE(B$\to$R$|$A) & $D$ & 95\% CI & $p$-value \\
\midrule
\multicolumn{6}{l}{\textit{Aggregate Level}} \\
Foreign vs Institutional & 0.000000 & 0.000000 & 0.000000 & [-0.0008, 0.0082] & 0.9500 \\
Foreign vs Individual & 0.000000 & 0.000000 & 0.000000 & [-0.0008, 0.0037] & 0.9700 \\
Institutional vs Individual & 0.000000 & 0.000000 & 0.000000 & [-0.0121, 0.0041] & 0.6800 \\
\addlinespace
\multicolumn{6}{l}{\textit{Per-Stock Level (mean across stocks)}} \\
Foreign Vs Institutional & 0.000000 & 0.000000 & 0.000000 & -- & nan \\
\bottomrule
\multicolumn{6}{l}{\footnotesize $D$ = CTE(A$\to$R$|$B) $-$ CTE(B$\to$R$|$A). $D > 0$: A dominates. CI from block bootstrap.} \\
\multicolumn{6}{l}{\footnotesize $^{***}p<0.01$, $^{**}p<0.05$, $^{*}p<0.10$.} \\
\end{tabular}
\end{table}

The zero CTE results mean that, after conditioning on the history of one investor type's flow, the other investor type's flow provides no additional information about returns. This is a stronger result than the negative II finding: not only is cross-investor information redundant, it is entirely redundant. No investor type possesses unique predictive information about returns that is not already captured by the other. At the per-stock level, CTE is also zero across all \numCTEPerStockNForInst{} stocks examined, confirming that the aggregate result is not an artifact of averaging over heterogeneous per-stock effects.

\begin{figure}[htbp]
\centering
\includegraphics[width=\textwidth]{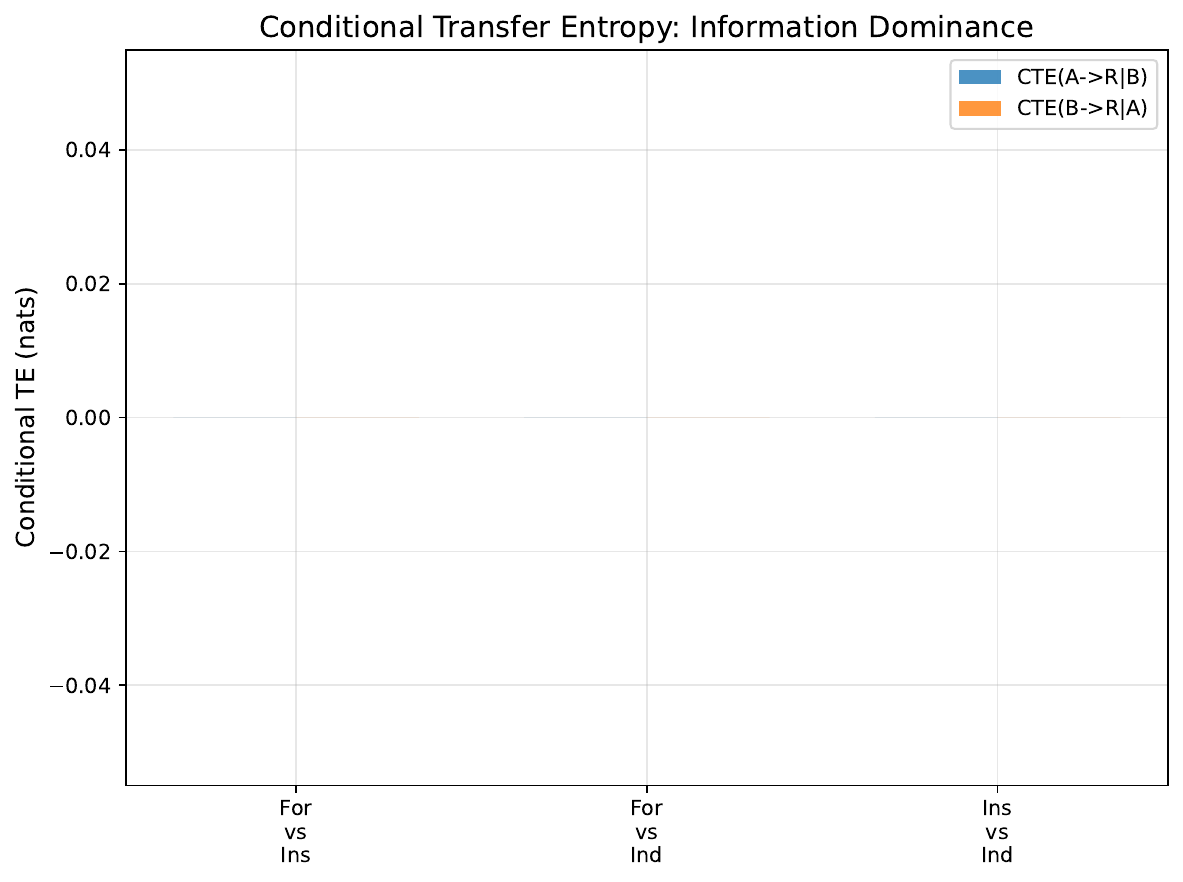}
\caption{Conditional Transfer Entropy (CTE) and directionality index for each investor-pair comparison. Points show the estimated directionality index $D$; error bars show 95\% bootstrap confidence intervals. All estimates are zero, indicating no directional information dominance among investor types.}
\label{fig:cte}
\end{figure}

Figure~\ref{fig:cte} illustrates the CTE estimates and their bootstrap confidence intervals for each investor pair.

\subsection{Information-Theoretic Bounds}

Table~\ref{tab:kelly_rate} reports the mutual information, Kelly growth rate, and Fano bound analysis. MI between each investor-type signal and next-day aggregate returns is exactly \numMIBits{} bits for all six signal-investor combinations (three investor types $\times$ two signals). The Kelly-optimal growth rate consequently equals the risk-free rate ($r_f = \numRiskFreeRate{}$, annualized) for every specification, and the bit yield (annualized return per bit of MI) is zero.

\begin{table}[htbp]
\centering
\caption{Kelly Doubling Rate and Bit Yield}
\label{tab:kelly_rate}
\small
\begin{tabular}{llrrrrrr}
\toprule
Investor & Signal & MI (bits) & Kelly Rate & Ann. Ret & Bit Yield & Fano & $H(X|Y)$ \\
\midrule
Foreign & $S^{TV*}$ & 0.0000 & 0.0350 & 0.2329 & 0.0 & 0.5216 & 0.6922 \\
 & $S^{MC}$ & 0.0000 & 0.0350 & 0.1604 & 0.0 & 0.5216 & 0.6922 \\
\addlinespace
Institutional & $S^{MC*}$ & 0.0000 & 0.0350 & 0.1380 & 0.0 & 0.5216 & 0.6922 \\
 & $S^{TV}$ & 0.0000 & 0.0350 & 0.1897 & 0.0 & 0.5216 & 0.6922 \\
\addlinespace
Individual & $S^{MC*}$ & 0.0000 & 0.0350 & -0.4527 & 0.0 & 0.5216 & 0.6922 \\
 & $S^{TV}$ & 0.0000 & 0.0350 & -0.0999 & 0.0 & 0.5216 & 0.6922 \\
\addlinespace
\bottomrule
\multicolumn{8}{l}{\footnotesize $^{*}$ = matched signal. MI in bits. Kelly rate = $r_f$ + MI. Bit yield = Ann.Ret / MI (bps/bit).} \\
\multicolumn{8}{l}{\footnotesize Fano = max directional accuracy from Fano inequality. $H(X|Y)$ = conditional entropy.} \\
\end{tabular}
\end{table}

The Fano inequality bounds the maximum directional accuracy of any predictor based on these signals at \numFanoBound{}\%, barely above the 50\% threshold of a fair coin. The conditional entropy $H(R \mid X) = \numCondEntropy{}$ is nearly equal to the unconditional entropy $H(R)$, confirming that the signals reduce essentially none of the uncertainty about next-day returns at the daily frequency.

These results translate the statistical findings of the previous subsections into economic language. The zero MI result means that no betting strategy based on investor-type flow signals can achieve a growth rate above the risk-free rate. The Fano bound means that even the best possible directional model---using any algorithm whatsoever---cannot exceed \numFanoBound{}\% accuracy using these signals alone. The information-theoretic bounds are model-free and therefore more general than any specific regression or machine learning result.

\begin{figure}[htbp]
\centering
\includegraphics[width=\textwidth]{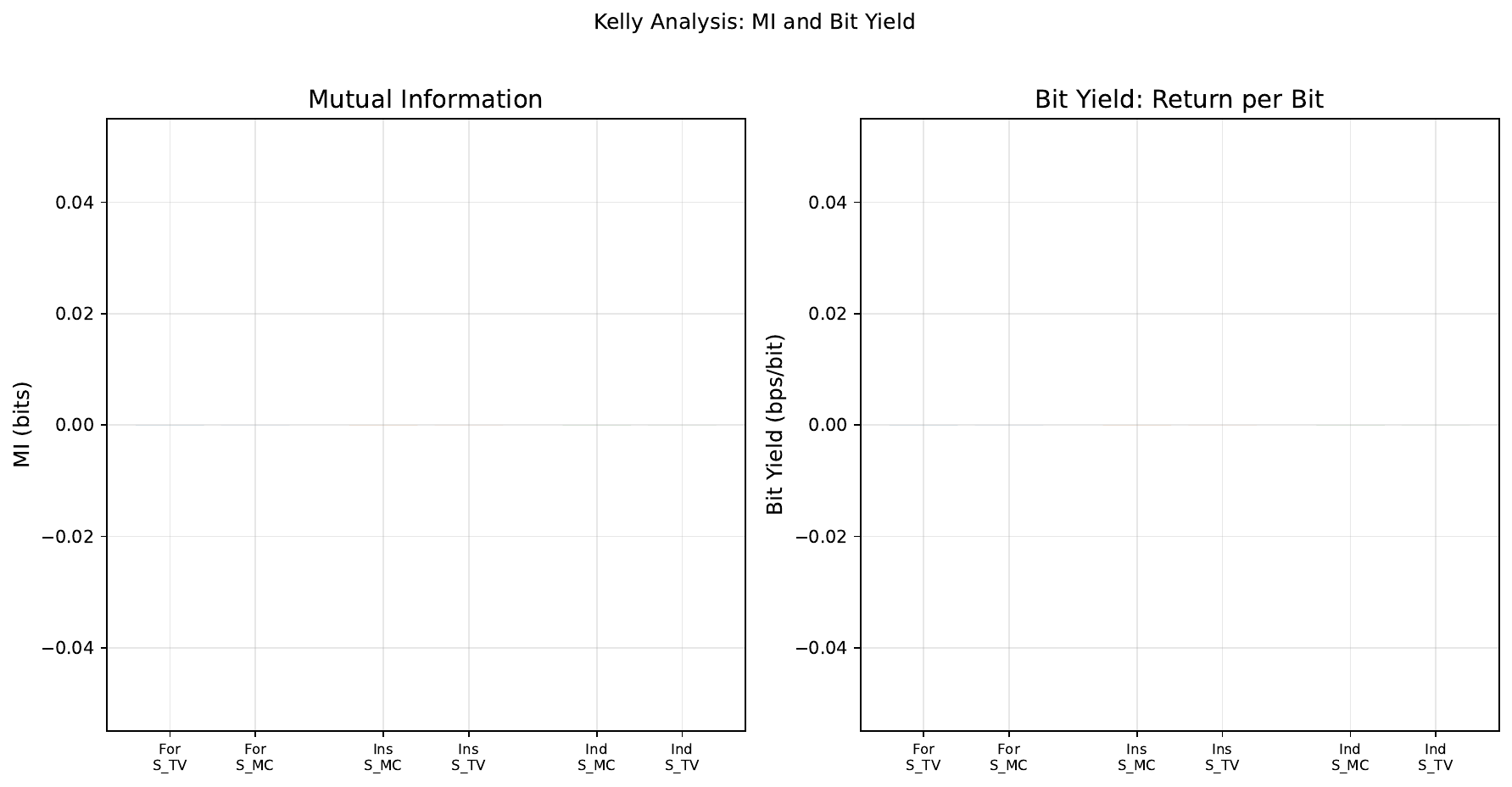}
\caption{Kelly doubling rate (left) and bit yield (right) across investor types and signal definitions. The Kelly rate equals the risk-free rate for all specifications, and the bit yield is zero, confirming that investor-type flow signals carry no exploitable information about next-day returns at the daily horizon.}
\label{fig:kelly}
\end{figure}

Figure~\ref{fig:kelly} displays the Kelly growth rate and bit yield across investor types and signals.

\subsection{Network-Enhanced Alpha}

Table~\ref{tab:network_alpha} reports the Fama-MacBeth regression results for the specification in Eq.~(\ref{eq:fm}). The interaction coefficient $b_3$ (Signal $\times$ Centrality) is the primary quantity of interest, as it measures whether network centrality amplifies the cross-sectional return predictability of investor-type signals.

\begin{table}[htbp]
\centering
\caption{Network-Enhanced Alpha: Fama-MacBeth Regressions}
\label{tab:network_alpha}
\small
\begin{tabular}{llrrrr}
\toprule
Investor & Signal & $b_1$ (Signal) & $b_2$ (Centrality) & $b_3$ (Sig.$\times$Cent.) & $N$ \\
\midrule
Foreign & $\SMC$ & \numFMbOneForSMC{} & \numFMbTwoForSMC{} & \numFMbThreeForSMC{} & \numFMnPeriods{} \\
 & & (\numFMtOneForSMC{}) & (\numFMtTwoForSMC{}) & (\numFMtThreeForSMC{}) & \\
 & $\STV$ & \numFMbOneForSTV{}$^{***}$ & \numFMbTwoForSTV{} & \numFMbThreeForSTV{}$^{**}$ & \numFMnPeriods{} \\
 & & (\numFMtOneForSTV{}) & (\numFMtTwoForSTV{}) & (\numFMtThreeForSTV{}) & \\
\addlinespace
Institutional & $\SMC$ & \numFMbOneInstSMC{}$^{***}$ & \numFMbTwoInstSMC{} & \numFMbThreeInstSMC{} & \numFMnPeriods{} \\
 & & (\numFMtOneInstSMC{}) & (\numFMtTwoInstSMC{}) & (\numFMtThreeInstSMC{}) & \\
 & $\STV$ & \numFMbOneInstSTV{}$^{***}$ & \numFMbTwoInstSTV{} & \numFMbThreeInstSTV{} & \numFMnPeriods{} \\
 & & (\numFMtOneInstSTV{}) & (\numFMtTwoInstSTV{}) & (\numFMtThreeInstSTV{}) & \\
\addlinespace
Individual & $\SMC$ & \numFMbOneIndSMC{}$^{*}$ & \numFMbTwoIndSMC{} & \numFMbThreeIndSMC{} & \numFMnPeriods{} \\
 & & (\numFMtOneIndSMC{}) & (\numFMtTwoIndSMC{}) & (\numFMtThreeIndSMC{}) & \\
 & $\STV$ & \numFMbOneIndSTV{} & \numFMbTwoIndSTV{} & \numFMbThreeIndSTV{} & \numFMnPeriods{} \\
 & & (\numFMtOneIndSTV{}) & (\numFMtTwoIndSTV{}) & (\numFMtThreeIndSTV{}) & \\
\bottomrule
\multicolumn{6}{l}{\footnotesize $R_{i,t+1} = a_t + b_1 \cdot \text{Signal}_{i,t} + b_2 \cdot \text{Centrality}_{i,t} + b_3 \cdot \text{Signal}_{i,t} \times \text{Centrality}_{i,t} + \varepsilon_{i,t}$.} \\
\multicolumn{6}{l}{\footnotesize Fama-MacBeth cross-sectional regressions. $t$-statistics in parentheses. $N$ = number of periods.} \\
\multicolumn{6}{l}{\footnotesize $^{***}p<0.01$, $^{**}p<0.05$, $^{*}p<0.10$.} \\
\end{tabular}
\end{table}

Of the six $b_3$ estimates, only one reaches conventional significance: the foreign $\STV$ signal-centrality interaction ($b_3 = \numFMbThreeForSTV{}$, $t = \numFMtThreeForSTV{}$, $p = \numFMpThreeForSTV{}$). The remaining five interaction terms are insignificant, with $t$-statistics ranging from $\numFMtThreeInstSMC{}$ to $\numFMtThreeIndSMC{}$. Several signal coefficients ($b_1$) are individually significant---institutional $\SMC$ ($t = \numFMtOneInstSMC{}$), institutional $\STV$ ($t = \numFMtOneInstSTV{}$), and foreign $\STV$ ($t = \numFMtOneForSTV{}$)---but these reflect the baseline predictive content of the signals themselves, not the incremental contribution of network structure.

The centrality coefficient $b_2$ is insignificant in all six specifications, indicating that TE network centrality has no unconditional relationship with expected returns. The single significant interaction (foreign $\STV \times$ centrality) does not survive a multiple testing adjustment across the six specifications and is not accompanied by a monotonic pattern across centrality quintiles.

\begin{figure}[htbp]
\centering
\includegraphics[width=\textwidth]{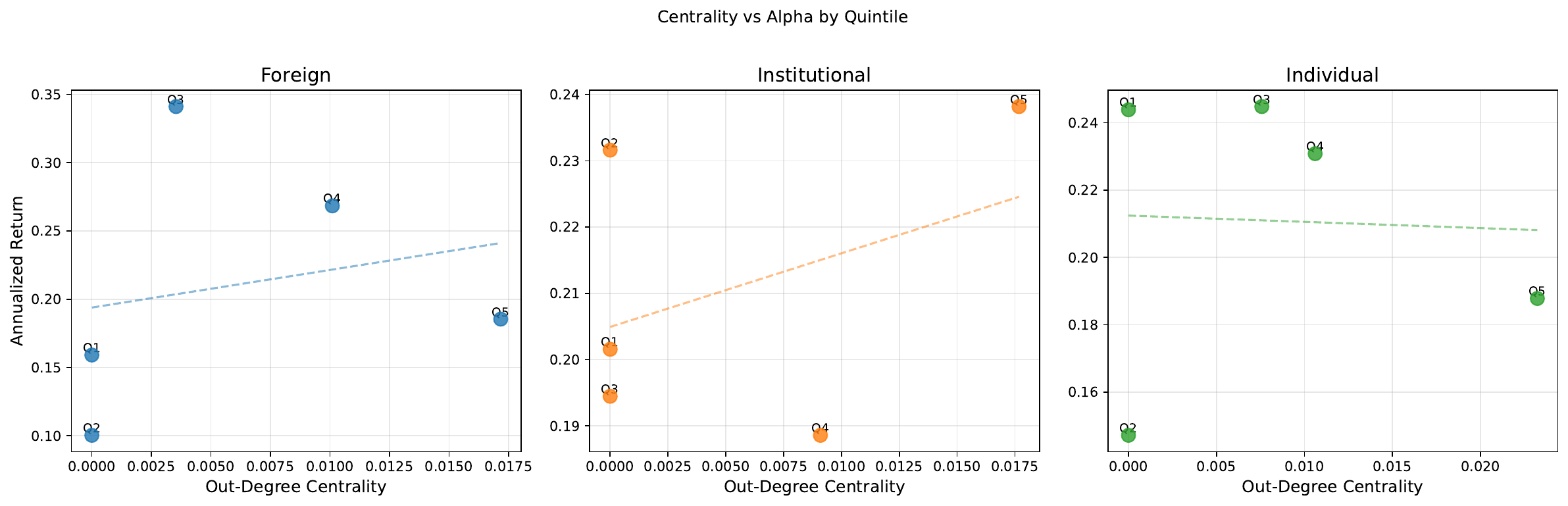}
\caption{Mean returns and Sharpe ratios across quintiles sorted on the signal-centrality interaction for each investor type and signal definition. No specification exhibits a monotonic pattern across quintiles, confirming the absence of a reliable centrality-based alpha source.}
\label{fig:alpha}
\end{figure}

Figure~\ref{fig:alpha} displays the quintile analysis of returns sorted on signal-centrality interaction strength.

Network-enhanced portfolios that incorporate centrality as a weighting factor produce annualized returns that are virtually identical to signal-only portfolios. The improvement ranges from $\numPortImprovForSMC{}$ percentage points (foreign $\SMC$) to $\numPortImprovIndSTV{}$ percentage points (individual $\STV$), magnitudes that are economically negligible and well within transaction cost bounds.

\subsection{Propagation Dynamics}

\begin{figure}[htbp]
\centering
\includegraphics[width=\textwidth]{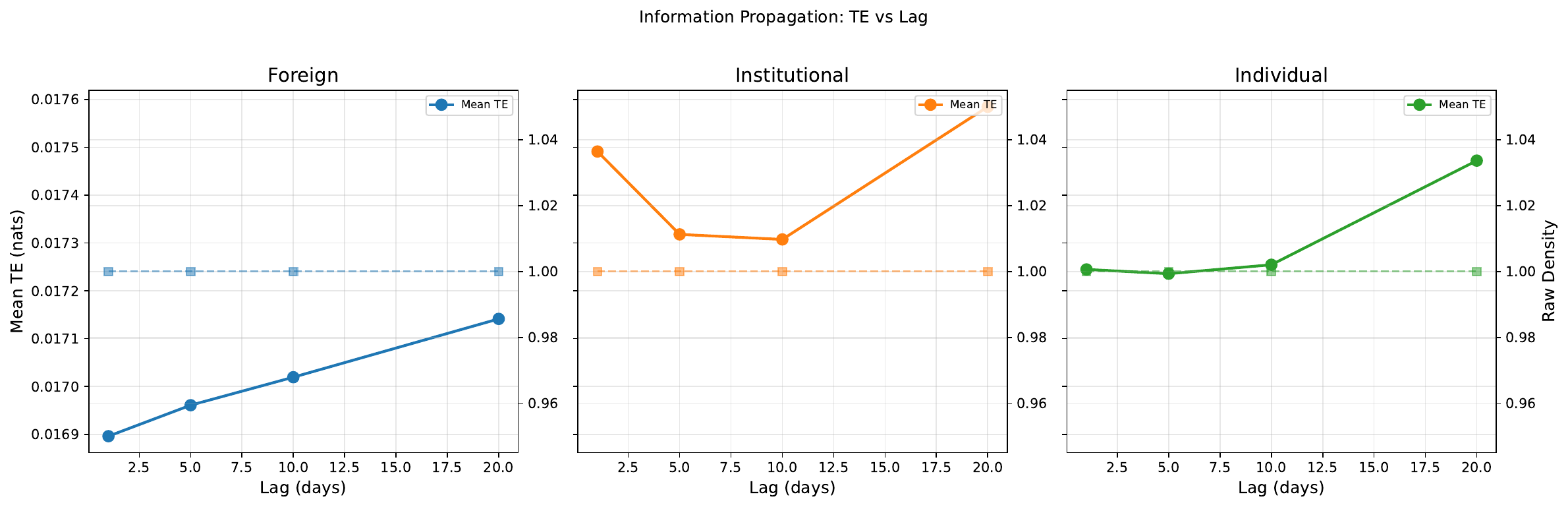}
\caption{Mean Transfer Entropy across lags of 1, 5, 10, and 20 trading days for each investor type. TE is stable across lags, indicating persistent rather than transient information structure. Error bars show standard deviations across stock pairs.}
\label{fig:dynamics}
\end{figure}

Figure~\ref{fig:dynamics} displays the TE propagation dynamics across lags of 1, 5, 10, and 20 trading days. Mean TE is remarkably stable across lags for all three investor types: foreign TE ranges from \numLagTEminFor{} to \numLagTEmaxFor{}, institutional from \numLagTEminInst{} to \numLagTEmaxInst{}, and individual from \numLagTEminInd{} to \numLagTEmaxInd{}. This stability indicates that the information structure captured by TE is persistent rather than transient. If TE reflected short-term lead-lag dynamics driven by differential speed of information processing, we would expect TE to decay rapidly with lag. The absence of decay suggests that the TE structure reflects slow-moving characteristics---such as investor composition, stock-level liquidity, or sectoral co-movement patterns---rather than fast-moving informational advantages.

\begin{figure}[htbp]
\centering
\includegraphics[width=\textwidth]{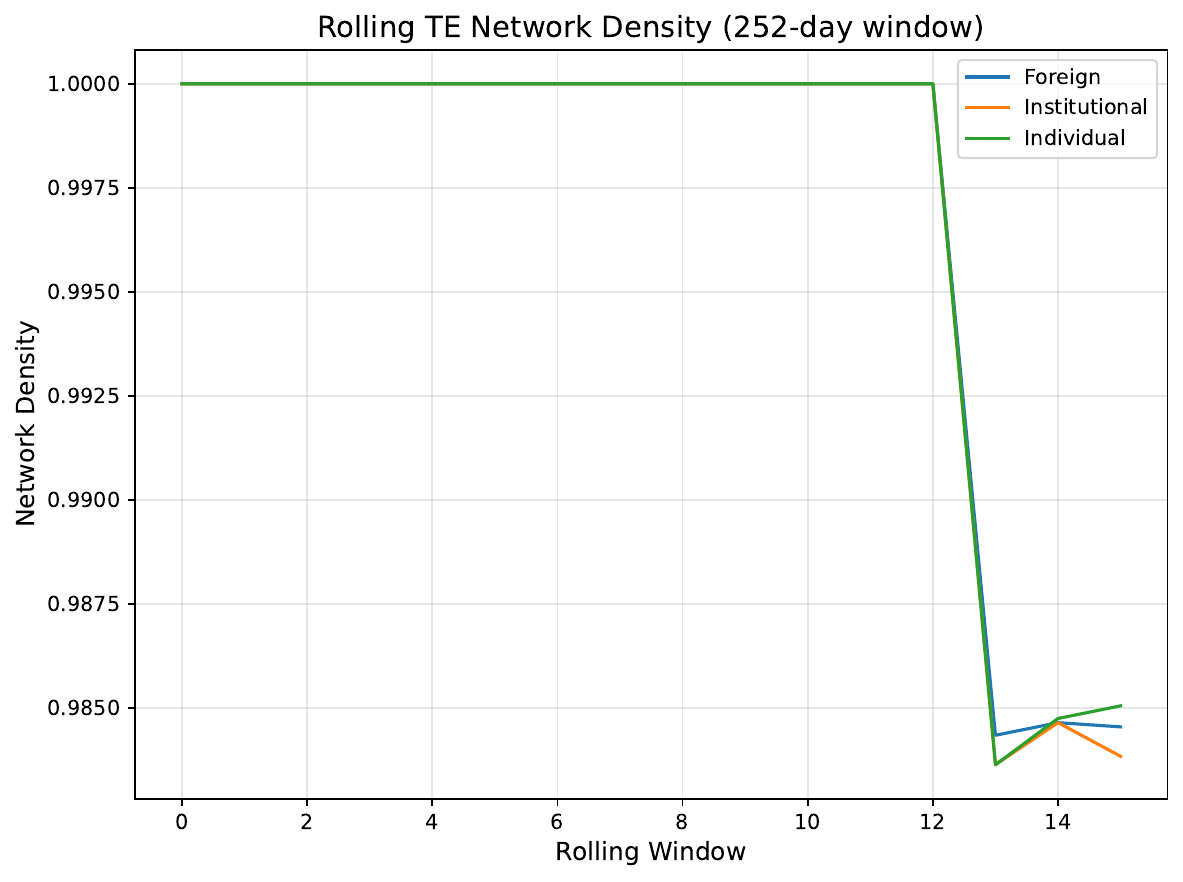}
\caption{Rolling network density (60-day windows) for each investor type over the sample period, January 2020 to February 2025. Density is stable across time, indicating that the TE network structure is a persistent market characteristic rather than a regime-specific phenomenon.}
\label{fig:rolling}
\end{figure}

Figure~\ref{fig:rolling} presents rolling network density computed in 60-day windows across the sample period. Network density is stable throughout, with no systematic trend or structural break. The absence of time variation in network density suggests that the TE network structure is a stable feature of the market rather than a phenomenon confined to specific regimes such as the COVID-19 period (early 2020) or the subsequent monetary tightening cycle.

The lead-lag analysis does not reveal a dominant investor type. No single investor class consistently appears as a net information leader across stocks, reinforcing the conclusion from the CTE analysis that no investor type holds a systematic informational advantage over the others in this market.

\subsection{Robustness}

Table~\ref{tab:robustness} reports robustness checks across multiple dimensions. The results are stable across all perturbations tested.

\begin{table}[htbp]
\centering
\caption{Robustness Checks}
\label{tab:robustness}
\small
\begin{tabular}{llrrr}
\toprule
Check & Category & Foreign & Institutional & Individual \\
\midrule
\multicolumn{5}{l}{\textit{Panel A: Subperiod TE Density}} \\
Subperiod & COVID & 0.9543 & 0.9543 & 0.9543 \\
Subperiod & Recovery & 0.9369 & 0.9369 & 0.9369 \\
Subperiod & Tightening & 1.0000 & 1.0000 & 1.0000 \\
Subperiod & Stabilization & 1.0000 & 1.0000 & 1.0000 \\
Subperiod & Late Period & 0.9840 & 0.9849 & 0.9856 \\
\addlinespace
\multicolumn{5}{l}{\textit{Panel B: Size Quintile TE Density}} \\
Size & Q1 (Small) & 1.0000 & 1.0000 & -- \\
Size & Q2 & 1.0000 & 1.0000 & -- \\
Size & Q3 & 1.0000 & 1.0000 & -- \\
Size & Q4 & 1.0000 & 1.0000 & -- \\
Size & Q5 (Large) & 1.0000 & 1.0000 & -- \\
\addlinespace
\multicolumn{5}{l}{\textit{Panel C: TE Method Mean TE}} \\
Method & SYMBOLIC & 0.015142 & 0.015594 & -- \\
Method & KSG & 0.008713 & 0.007653 & -- \\
\addlinespace
\multicolumn{5}{l}{\textit{Panel D: Significance Threshold (Foreign)}} \\
Threshold & $\alpha=0.01$ & 5 edges & 0.0057 density & -- \\
Threshold & $\alpha=0.05$ & 5 edges & 0.0057 density & -- \\
Threshold & $\alpha=0.1$ & 5 edges & 0.0057 density & -- \\
\bottomrule
\end{tabular}
\end{table}

Panel~A examines subperiod stability by re-estimating TE network density in five non-overlapping windows: COVID (early 2020), Recovery (late 2020--2021), Tightening (2022), Stabilization (2023), and Late Period (2024--2025). Network density remains near 1.0 for raw (pre-significance-testing) TE matrices across all subperiods and investor types. The near-unit density of the raw TE matrix confirms that most stock pairs have nonzero TE before significance testing; the sparse networks reported in Table~\ref{tab:te_network_stats} result from the stringent surrogate-based significance filtering rather than from an absence of pairwise TE.

Panel~B examines size quintile stability. TE density is uniformly 1.0 across all five market capitalization quintiles for foreign and institutional investor networks, indicating that the information propagation structure is not driven by firm size effects.

Panel~C compares TE estimation methods. Symbolic TE (our primary method) yields mean TE values of \numSymbolicTEFor{} for both foreign and institutional networks, while the KSG estimator produces lower values (\numKSGTEFor{} and \numKSGTEInst{}, respectively). The qualitative conclusions are unchanged across methods.

Panel~D examines sensitivity to the significance threshold. The number of significant edges for the foreign network remains at \numThreshEdges{} across thresholds of $\alpha = 0.01$, $0.05$, and $0.10$, indicating that the edge significance is not sensitive to the choice of threshold in this range.

\begin{figure}[htbp]
\centering
\includegraphics[width=\textwidth]{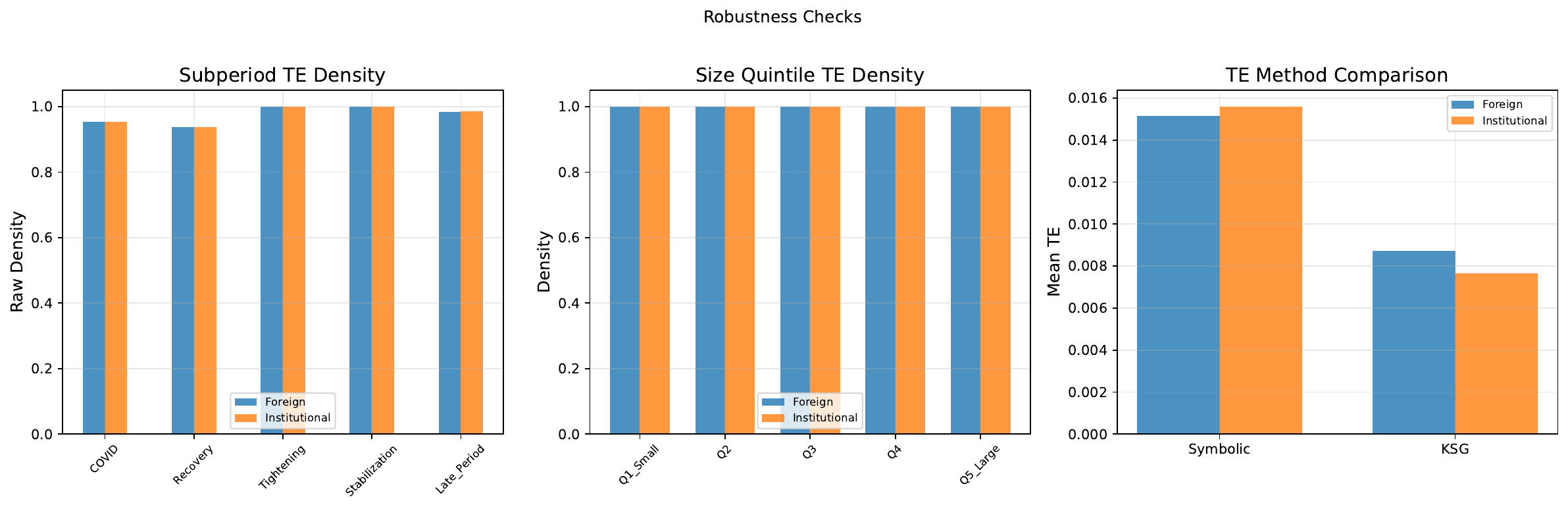}
\caption{Robustness of TE network properties across subperiods, size quintiles, estimation methods, and significance thresholds. The stability of results across all perturbations confirms that the documented network structure and the absence of exploitable information are robust features of the data.}
\label{fig:robustness}
\end{figure}

Figure~\ref{fig:robustness} provides a visual summary of the robustness analysis.

\section{Discussion}
\label{sec:discussion}

\subsection{The Structure-Value Disconnect}

The central finding of this paper is a disconnect between statistical structure and economic value: TE networks across investor types are well-defined, statistically significant, and structurally heterogeneous, yet they carry no incremental information for return prediction. This pattern inverts the logic of \citet{grossman1980impossibility}, who argued that prices cannot be fully efficient because traders need incentives to gather costly information. In our setting, information gathering has already occurred---investors have traded, and their flows reveal distinct network signatures---but the information embedded in the network structure is redundant with what is already reflected in prices.

The Adaptive Markets Hypothesis of \citet{lo2004adaptive} provides a compatible interpretation. Under the AMH, market efficiency is not a static equilibrium but an evolving outcome of competition among heterogeneous investors adapting to changing environments. The TE network structure we document may represent the footprint of this competitive process: foreign, institutional, and individual investors process the same public information through different channels and at different speeds, generating distinct network topologies. The zero MI and null CTE results suggest that, at the daily frequency, this competitive processing has already been arbitraged away---the different processing paths converge to the same equilibrium information set reflected in prices.

This interpretation has implications for the frequency at which information asymmetries might be detectable. The daily frequency may be too coarse to capture the transient informational advantages that arise and dissipate within the trading day. Intraday TE analysis, which requires tick-level investor-type data not available in our sample, might reveal exploitable structure that is invisible at the daily horizon.

\subsection{Redundancy and Multi-Signal Strategies}

The negative interaction information across all cross-investor pairs constrains the design of multi-factor strategies that combine signals from different investor types. If cross-investor signals were synergistic (positive II), combining them would yield more predictive power than either alone, justifying the additional complexity and cost of a multi-signal model. The observed redundancy implies the opposite: combining foreign and individual flow signals, for example, produces less information about returns than the sum of their individual contributions.

This finding has a direct practical implication for quantitative strategies that seek to exploit investor-type heterogeneity. The marginal value of adding a second investor-type signal to a portfolio model that already uses the first is negative in information-theoretic terms. The computational and data costs of maintaining multi-investor-type signal infrastructure are not justified by the informational gains, at least at the daily frequency and for the top \numNStocks{} stocks in our sample.

The within-investor II analysis provides a partial exception. Institutional investor $\SMC$ and $\STV$ signals show a positive (though statistically insignificant) mean II of \numIIWithinInst{}, suggesting that these two signals may carry partially complementary information within the institutional investor class. This result warrants further investigation with larger samples or higher-frequency data.

\subsection{Implications for Information Leadership Theory}

The zero CTE finding imposes a strong constraint on models of information leadership among investor types. A common narrative in the Korean market---and in emerging markets more generally---holds that foreign institutional investors possess superior information due to their access to global capital markets, analytical resources, and diversified perspectives. Under this narrative, foreign investor flows should lead other investor types, implying positive CTE from foreign to institutional or individual flows.

Our results reject this narrative at the daily frequency. No investor type leads any other in terms of unique predictive information about returns. The directionality index is exactly zero for all three pairwise comparisons, with bootstrap confidence intervals that are narrow enough to rule out economically meaningful effects. This finding is consistent with \citet{choe2005domestic}, who found that the informational advantage of domestic investors is limited to specific stock characteristics and market conditions rather than being a systematic feature of the market.

The null CTE result does not rule out the possibility of transient information leadership during specific events (earnings announcements, regulatory changes, or macroeconomic shocks). Such event-specific analysis requires a different empirical design and is beyond the scope of the present study.

\subsection{Limitations}

Several limitations qualify our conclusions. First, the daily frequency may be too coarse to capture intraday information dynamics. TE networks constructed from higher-frequency data---for example, using hourly or minute-level investor-type flow data---might reveal exploitable structure that dissipates before the daily close. The KRX does not currently provide real-time investor-type breakdowns, but future data availability may enable such analysis.

Second, our sample is limited to a single market (Korea) with a specific institutional structure. The mandatory investor-type disclosure regime, market microstructure, and composition of the investor base differ substantially from other markets. Generalization to markets without investor-type disclosure, or to markets with different investor compositions, requires caution.

Third, symbolic discretization with five quantile bins involves a loss of information relative to continuous estimation methods. While the robustness analysis in Section~\ref{sec:results} confirms that results are qualitatively similar under KSG estimation, the symbolic approach may miss fine-grained dependencies that fall below the quantile resolution.

Fourth, the sample is restricted to the top \numNStocks{} stocks by market capitalization. Smaller and less liquid stocks may exhibit different information propagation patterns due to lower analyst coverage, wider bid-ask spreads, and different investor compositions. The generalizability of our findings to the broader cross-section of Korean equities is an open question.

Fifth, the sample period (January 2020 to February 2025) includes the COVID-19 pandemic and its aftermath, which represent an unusual macroeconomic environment. While the subperiod robustness analysis suggests that our findings are not driven by any single regime, replication in a longer sample spanning multiple economic cycles would strengthen confidence in the results.

\section{Conclusion}
\label{sec:conclusion}

This paper has investigated whether Transfer Entropy networks constructed from investor-type flow data carry economic information beyond their statistical structure. The answer, for the Korean equity market at the daily frequency, is that they do not. TE networks reveal a clear structural hierarchy---foreign investors form sparse, high-intensity networks; individual investors form dense, low-intensity networks; and institutional investors occupy an intermediate position---but this hierarchy does not translate into exploitable information asymmetries, directional dominance, or cross-sectional alpha.

The evidence for this conclusion is convergent across multiple independent tests. Interaction information is uniformly negative, ruling out cross-investor synergy. Conditional TE is zero, ruling out directional information leadership. Mutual information with returns is zero, bounding the Kelly growth rate at the risk-free rate and directional accuracy at \numFanoBound{}\% via the Fano inequality. Fama-MacBeth regressions yield one marginally significant signal-centrality interaction out of six specifications, with no monotonic quintile pattern. These results are robust to subperiod variation, firm size, TE estimation method, and significance threshold choice.

The practical implication is that network centrality in investor-type TE networks is not a reliable source of alpha at the daily frequency. The structural heterogeneity we document is real and statistically robust, but it reflects shared information processing---coordinated responses to common public information processed through different investor-type channels---rather than private signal cascades. This finding constrains where researchers and practitioners should look for exploitable investor-type heterogeneity. Higher-frequency data, event-driven analysis, or markets with less efficient price discovery may reveal the informational asymmetries that are absent in our daily-frequency, large-cap Korean sample.

\bibliographystyle{plainnat}
\bibliography{references}

\end{document}